# Classification of Autism Spectrum Disorder Using Supervised Learning of Brain Connectivity Measures Extracted from Synchrostates

Wasifa Jamal[1], Saptarshi Das[1], Ioana-Anastasia Oprescu[1], Koushik Maharatna[1], Fabio Apicella[2], and Federico Sicca[2]

1) School of Electronics and Computer Science, University of Southampton, Southampton SO17 1BJ, United Kingdom.
2) IRCCS Stella Maris Foundation, Pisa (Calambrone), Italy.

**Authors' Emails:**

wj4g08@ecs.soton.ac.uk (W. Jamal*)

sd2a11@ecs.soton.ac.uk, s.das@soton.ac.uk (S. Das)

io1g10@ecs.soton.ac.uk (I. Oprescu)

km3@ecs.soton.ac.uk (K. Maharatna)

fabio.apicella@fsm.unipi.it (F. Apicella)

fsicca@fsm.unipi.it (F. Sicca)

**Corresponding author's phone number:** +44(0)7508408746

**Abstract:**

*Objective*. The paper investigates the presence of autism using the functional brain connectivity measures derived from electro-encephalogram (EEG) of children during face perception tasks. *Approach*. Phase synchronized patterns from 128-channel EEG signals are obtained for typical children and children with autism spectrum disorder (ASD). The phase synchronized states or *synchrostates* temporally switch amongst themselves as an underlying process for the completion of a particular cognitive task. We used 12 subjects in each group (ASD and typical) for analyzing their EEG while processing fearful, happy and neutral faces. The minimal and maximally occurring synchrostates for each subject are chosen for extraction of brain connectivity features, which are used for classification between these two groups of subjects. Among different supervised learning techniques, we here explored the discriminant analysis and support vector machine both with polynomial kernels for the classification task. *Main results*. The leave one out cross-validation of the classification algorithm gives 94.7% accuracy as the best performance with corresponding sensitivity and specificity values as 85.7% and 100% respectively. *Significance*. The proposed method gives high classification accuracies and outperforms other contemporary research results. The effectiveness of the proposed method for classification of autistic and typical children suggests the possibility of using it on a larger population to validate it for clinical practice.

## 1. Introduction

The human brain is a complex biological organ. Neural assemblies in the brain synchronize and form functional associations which can be configured into a network. These networks share features with other networks from biological and physical systems and so inherently can be classified as complex networks. Thus the standard network characterization techniques of the complex networks can be applied for reliable and effective quantification of the brain connectivity graphs. The comparison of brain network topologies between subject populations can reveal presumed abnormalities and may also lead to the identification of distinguishable measures for the classification of such populations [1].

Autism is a lifetime condition which Minshew and William defined as a polygenetic developmental and neurobiological disorder [2] that is characterized by atypical behavior and lack of social reciprocity. ASD comprises a complex group of behaviorally defined conditions with core deficits in social interaction, communication and stereotyped and restricted behaviors. Although problems in perception and attention are not considered as primary diagnostic categories, individuals with ASD often present difficulties in these domains. It is believed these constellations suggest generalized dysfunction in the association cortex. Researchers have suggested that autism is due to under functioning integrative circuitry leading to deficits in neural level information integration [3], [4]. Research has shown that a key feature in the neuro-anatomy of autism is the early brain overgrowth [5] subsequently leading to greater local connectivity and suppressed long-range connectivity. Findings from [6–9] show evidence of overgrowth of shorter range cortico-cortical intra-hemispheric connections with little involvement of connections between hemispheres and cortex and subcortical structures. The behavioral symptoms of autism could be a manifestation of these disrupted neural circuits. Evidence has been found for supporting the hypothesis of under-connectivity within large-distant networks and also under-connectivity within the local networks. These evidences come from several studies done using functional Magnetic Resonance Imaging (fMRI), EEG and Magneto-encephalogram (MEG) recording. Kleinhans *et al.* [10] found disrupted functional connectivity between the Fusiform face area and the left amygdala and between the posterior cingulate and thalamus in the social brain during face processing. An fMRI based study carried out by Just *et al.* reported evidences of overall functional under-connectivity in autistic subjects compared to controls [3]. Tommerdahl *et al.* [11] detected local under-connectivity in autistic adults in their study of sensory perception. Coben *et al.* [12] found reduced inter-hemispheric coherence across different frequency bands. Bosl *et al.* used the complexity of EEG signals recorded during resting state as a feature to distinguish typically developing children from children with the risk for ASD [13]. It is very important to understand how these finding translate into differences in the functional organization of the brain between typical and high risk populations. Thus studying the functional connectivity patterns may be the key to understanding the differences between the typical and autistic brain. The use of graph theory provides a new dimension to the investigation of the brain network organization in humans at different levels of granularity [14] and thus gives a holistic analysis. Complex network metrics have been used to model the organization of the human brain in various studies conducted on fMRI and MEG recoding. Supekar *et al.* [15] applied graph theory on task-free Alzheimer's disease fMRI. Liu *et al.* [16] investigated the network properties of functional networks of schizophrenic patients from resting state fMRI data, Bassett *et al.* [17] used similar graph metrics for MEG analysis of healthy adults during resting state and finger–tapping task. Studies in this field suggest that in complex networks more robust results are obtained by retaining weight information of the graphs as compared to the binary graphs [18] [19], since binary networks only provide an approximation of the original weighted network as the whole range of connection strengths are lost. It is also known that weighted characterization is useful to filter out the influence of weak and potentially less-significant links [1]. Therefore quantitative characterization of the connectivity derived from phase synchronization measures in ASD patients may lead to an effective classifier enabling intervention at appropriate stage. Here we use graph theory or complex weighted network measures as features to classify the typically developing and autistic children using different supervised learning based classification algorithms.

Recently there has been a lot of interest and ongoing research in the field of dynamic connectivity analysis owing to the inherent dynamic nature of the human brain using EEG, MEG and fMRI data. The works by Mutlu *et al.*[20], Betzel *et al.*[21] and Dimitriadis *et al.*[22] used popular techniques like phase locking value, correlation and phase lag index to estimate function connectivity to further investigate the nature of these connectivity maps. The work by Jamal *et al.* [23] established the concept of finding unique stable phase synchronization patterns over the scalp, from EEG acquired during face perception task, which is termed as synchrostates. The concept of synchrostate has been used to derive brain connectivity networks for autistic and typically growing population in [24]. The study in [24] shows that the respective modularity values of the population average functional networks for the two groups (ASD and typical) have notable differences. Subsequently, the paper proposes the use of modularity values of brain networks as a marker for distinguishing between the two classes. Here we extend the concept and analysed each individual subject's network metrics acquired after performing the synchrostate analysis in conjunction with the application of complex graph theoretic methods on the maximum (frequently) and minimum (rarely) occurring states. Since different subjects may need less/more number of optimum synchronised states for the same face-perception task, we restricted our study to graph measures of the maximum and minimum occurring states only.

Deficiencies in children with autism in understanding social information conveyed by emotional faces have been attributed to the inability in activation of brain circuitry involved in face processing [25]. This impairment in social processing is said to be a core difficulty in autism [26]. Thus we explore the brain connectivity of children in two populations to investigate if any discernible differences can emerge from information about the functional organization of their brain during face perception. We use the EEG recording from 12 ASD and 12 typical children during the execution of a face perception task to get the individual synchrostates and their transition sequence. Based on the frequency of occurrence for each state and their temporal stability, we form the connectivity maps and compute the graph metrics for the maximum and minimum occurring states for every individual for each of the three stimuli. The parameters are used as features and then a feature ranking algorithm is employed to sort the features according to their decreasing level of class-discrimination capability. After being ranked, grouped and arranged, the features are used to train a few supervised machine learning algorithms for classification. Next, the performance of each of the classifiers for different sets of features is compared to identify the combination of best feature subset and classifier.

The simple linear classifier is a very popular algorithms for event related potential (ERP) analysis in brain-computer interface (BCI) applications [27], [28] which have been applied for classification of motor imagination [29]. Garrett *et al.* [30] applied both linear discriminant analysis (LDA), conventional neural networks and SVM on EEG recorded during mental tasks and concluded SVM to be the more sound and conclusive algorithm, although the other two did not perform much worse than the other. Variants of the classical SVM learning algorithm have been applied to classification problems in EEG for application in BCI systems [31–33].In this work we not only focus on the performance of different classifiers to be used in our binary class classification problem but we also focus to find the optimal combination of features to improve the discrimination capability of the classes.

## 2. Method
### 2.1. Experimental data and EEG preprocessing

Data from a high-density EEG study was used for the current exploration. The experimental sample of the dataset contains EEG data from 24 participants; 12 children with ASD and 12 healthy controls. The ASD group had subjects with an age range of 6-13 years with a mean age of 10.2 years. The control population was aged between 6-13years as well with an average age of 9.7 years. Three kinds of emotional face stimuli were presented, displaying fearful, neutral and happy expressions. The experiment was conducted in 4 blocks and in each block 10 fearful, 10 neutral and 10 happy faces were presented at random. Data was acquired at 250 Hz using a 128-channel HydroCel Geodesic Sensor net [25].

The continuous EEG data was segmented into 1000ms epochs and segments with signals over a threshold of 200μV were rejected. Data was band-pass filtered from 0.5 Hz to 50 Hz and baseline corrected. Then the synchrostate derivation algorithm of Jamal *et al.* [23] was run on the data. For obtaining the synchrostates and there subsequent brain connectivity metrics, we need to apply the following steps: 1) define a measure to produce the time varying phase information amongst the EEG electrodes, 2) cluster the characteristic phase difference patterns and use the synchronization index to quantify the temporal stability of each synchrostate, and the 3) translate the unique clusters into a complex brain network using a graph theoretic approach and then derive quantitative measures for each connectivity map.

Phase synchronization has been determined as a characteristic for communication between the regions in the brain [34–36] acting as the key manifestation of the underlying mechanism of information exchange occurring during the execution of a task. With the objective of tracing the phase synchronization patterns in the EEG signal continuous wavelet transform (CWT) was applied to each channel data using a complex Morlet mother wavelet. The arguments of the transformed complex frequency domain representations are then used to compute the instantaneous phase for different EEG channels. Taking the difference between the instantaneous phases of two channels $i$ and $j$ yields the pairwise phase difference of the EEG signals as a function of time ($t$) and frequency or scale ($a$).

$$\Delta \varphi_{ij}(a,t) = \varphi_i(a,t) - \varphi_j(a,t) \quad (1)$$

This technique when applied across all channels generates a stream of cross channel phase difference matrices which are square and symmetric with zero diagonal elements as the diagonal represents the phase difference of an electrode to itself. At every time instant, the matrices were averaged across the trials for each stimulus. Observing the resultant multi-channel phase data in time reveals the existence of distinctive patterns which are stable for finite duration in the order of milliseconds. Thus the next step is to identify similar topographies over the time duration of interest. These stable patterns are optimally translated into synchrostates using the *k*-means clustering, to associate similar patterns into a single class [23]. The *k*-means clustering is a classical unsupervised learning pattern recognition technique, which uses Euclidean distance to measure the dissimilarity between data vectors thus grouping all the phase matrices with similar patterns. The algorithm iteratively minimizes the cost function $J(\theta, U)$ as shown in (2).

$$J(\theta, U) = \sum_{i=1}^{N} \sum_{j=1}^{m} u_{ij} \left\| x_i - \theta_j \right\|^2 \qquad (2)$$

where, $\theta = \begin{bmatrix} \theta_{T_1} & \cdots & \theta_{T_m} \end{bmatrix}^T$, $\|\cdot\|$ is the Euclidean distance, $\theta_j$ is the mean vector of the $j^{th}$ cluster and $u_{ij} = 1$, if $x_i$ lies closest to $\theta_j$; 0 otherwise [31]. In this case, $X$ is the dataset of all pairwise EEG instantaneous phase differences which is a function of time and frequency. We cluster $X$ along time (1 second duration for the current study), to find out unique phase synchronized patterns for a particular frequency band. The result gives $k$ number of centroids and a vector with the corresponding state labels for every time instance.

The typical behavior of synchrostates is that they start with an almost constant phase difference topology (few ms) and configuration remains stable for a finite duration of time. It then suddenly switches to another stable phase synchronized pattern and gets locked for some duration before it switches to another state or the previous one. The optimal number of synchrostates may vary for each individual and the inter-state switching patterns can be thought as a manifestation of the intricate processing steps within brain. It is found in [24] that some of states occur more frequently and others less frequently which are termed as maximum (max) and minimum (min) state hereafter in this paper. The topography of the max-state and min-state along with the inter-state switching characteristics for one typical and one ASD child have been shown in Figure 1 as an example. It shows a comparison of a typical and an autistic child along with the head topography of the average phase difference corresponding to the maximum and minimum occurring states. It is evident from Figure 1 that the state 2 and state 3 are the minimum occurring for typical and ASD cases respectively. Whereas, the maximum occurring states for these two cases are case 3 and case 1 respectively.

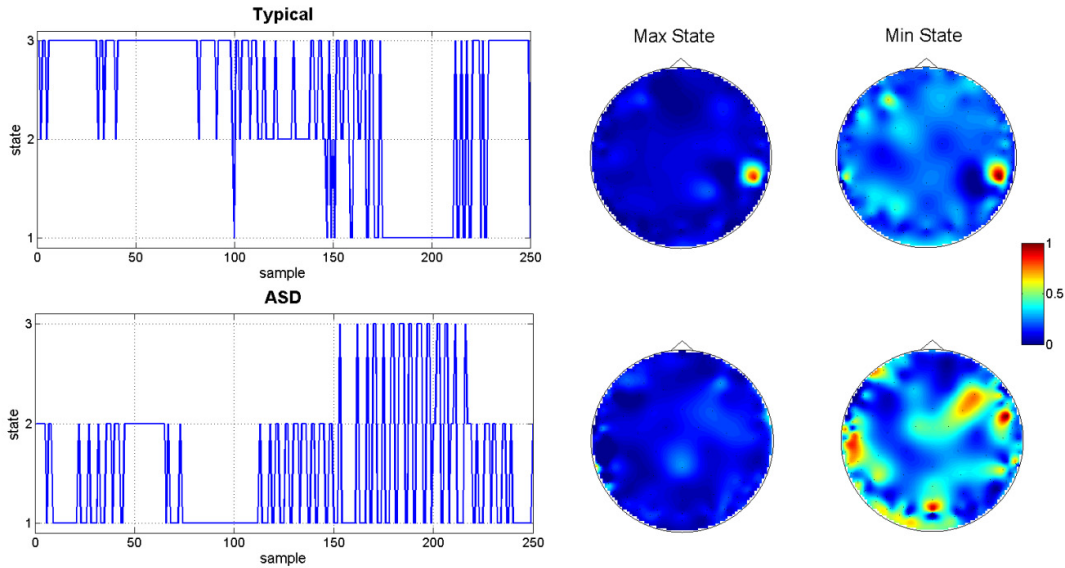

Figure 1: State switching time course and head plots of maximum and minimum occurring state for a typical and ASD child.

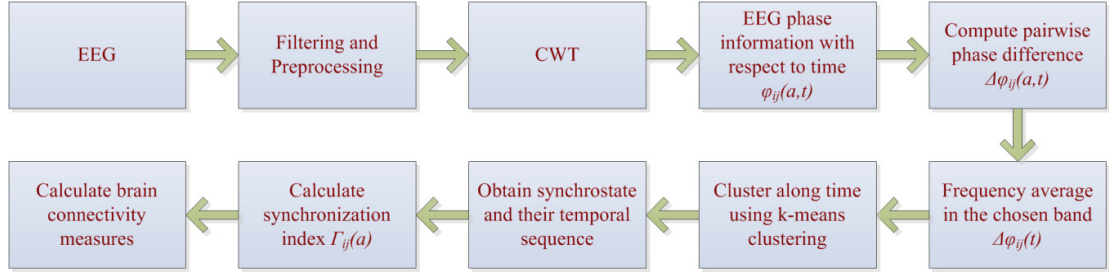

Figure 2: The processes involved in deriving synchrostates and brain connectivity measures.

Once these synchrostates are computed for every subject and all of the three stimuli, their transition patterns for the duration of a task in a specific band can be utilized to investigate the temporal evolution of phase synchronization and hence the functional connections relating to the task. Researchers have argued that gamma-band (30 Hz and above) synchronization is the key process that reflects underlying cortical computations [37]. Thus we compute gamma band phase synchronization in synchrostates to characterize the underlying connections that are formed in the autistic and typical brain during the execution of the face perception task. A phase synchronization index given in (3) is used to quantify the average temporal stability of the synchrostate in the gamma band.

$$\Gamma_{ij}(a) = \frac{1}{N}\sqrt{\left[\sum_t \cos(\Delta\varphi_{xy}(a,t))\right]^2 + \left[\sum_t \sin(\Delta\varphi_{xy}(a,t))\right]^2} \quad (3)$$

Here, $\Gamma_{ij}(a) \in [0,1]$ and $N$ is the number of time points for which the synchrostate occurs. Values of $\Gamma_{ij}(a)$ close to 1 indicate that the phase difference between the signals at the given wavelet scale vary little over time and therefore can be considered in synchrony. To get the gamma band synchronization we compute the average $\Gamma_{ij}$ for all the frequencies of gamma band. This index is calculated for each state and for each channel pairs resulting in a symmetric and square global synchronization matrix which describes the gamma band phase synchronization for each synchrostate in the entire EEG space [24].

Figure 2 shows flowchart for the whole process of deriving synchrostates from the EEG signals and subsequently obtain the corresponding brain connectivity measures.

*2.2. Feature selection*

The global synchronization matrix can now be translated into a complex network that is useful to shed light on the phase synchrony amongst different regions and hence describe the nature of the functional network configuration of the brain. The brain connectivity map can be configured by considering the EEG electrodes as nodes and the synchronization values between them as the weighted edges i.e. connection strength between the nodes. The appropriate graph metrics when studied can facilitate the interpretation of the topological properties and intrinsic meaning of the functional brain networks. The two types of generic measures that are most relevant in this particular study for understanding the autistic and typical brain's capability for information processing are segregation and integration. Measures of segregation in a brain network account for the ability of segregated specialized neural processing within highly connected brain regions. It has been used as the means to evaluate the local connectivity [1]. The common measures of integration are capable of capturing the capacity of global interaction in a network and estimate the ease of network–wide communication [38]. Features which give the most appropriate information regarding network formation were selected for the study. The chosen features have been listed in Table 1 along with the physical network attributes they portray.

These features were generated by averaging the phase response of the CWT across all channels for each individual over all trials corresponding to a single stimulus and carrying out the synchrostate analysis. From the resulting synchrostate sequence of each child the complex network parameters of the connectivity network are computed from the functional connectivity graphs of the maximum and minimum occurring states. We use the maximum and minimum states since previous study of synchrostates in autistic population [24] showed distinguishing properties in graph metrics derived from these states. The phase-locked matrices obtained from clustering, the maximum and minimum occurring synchrostates can be converted to analogous undirected graph using the synchronization index in (3). In the undirected brain network, each edge represents the value of the synchronization index as the coupling strength between two electrodes and has been represented in Figure 3 for a typical and ASD child.

Table 1: Different features of the brain connectivity graphs used for classification

| Name | Mathematical description | Physical significance |
|---|---|---|
| Modularity | $Q^w = \dfrac{1}{l^w} \sum_{i,j \in N} \left[ w_{ij} - \dfrac{k_i^w k_j^w}{l^w} \right] \delta_{m_i, m_j}$; <br> $w_{ij}$ is the connection weights, $k_i^w = \sum_{j \in N} w_{ij}$ is the weighted degree, $l^w = \sum_{i,j \in N} w_{ij}$ is the sum of all weights in the network. Also, $\delta_{m_i, m_j} = 1$ if $m_i = m_j$, and 0 otherwise ($m_i$ is the module containing node $i$) | Measure of segregation. Quantifies the degree to which a network can be subdivided into a group of nodes with small number of between group links (edges) and large number of within group links [1]. |
| Transitivity | $T^w = \dfrac{1}{n} \dfrac{\sum_{i \in N} 2 t_i^w}{\sum_{i \in N} k_i (k_i - 1)}$; <br> $t_i^w = \dfrac{1}{2} \sum_{j,h \in N} \left( w_{ij} w_{ih} w_{jh} \right)^{1/3}$ is the weighted geometric mean of the triangles around $i$ | It is the ratio of the triangle to triplets of the network. Is a measure of segregation in complex network analysis [39]. |
| Characteristic path length ($L^w$) | $L^w = \dfrac{1}{n} \sum_{i \in N} \dfrac{\sum_{j \in N, j \neq i} d_{ij}^w}{n - 1}$; <br> $d_{ij}^w$ is the shortest weighted path length between $i$ and | It is essentially the global mean of the distance matrix i.e. the average of the shortest path length between a node and all other nodes [40]. It is a measure of network integration. |

| Global efficiency ($E^w$) | $E^w = \dfrac{1}{n}\sum\limits_{i\in N}\dfrac{\sum\limits_{j\in N, j\neq i}\left(d_{ij}^w\right)^{-1}}{n-1}$ | A measure of integration, that is the calculated by averaging the inverse of the distance matrix [38]. |
|---|---|---|
| Radius | $e_i^w = \max\left(d_{ij}^w \circ d_{ij}^w\right), r^w = \min\left(e_i^w\right)$ | Radius is derived from a network's eccentricity ($e_i^w$) which refers to the maximum value of each row of the Hadamard (dot) product of $d_{ij}^w$ |
| Diameter | $e_i^w = \max\left(d_{ij}^w \circ d_{ij}^w\right), D^w = \max\left(e_i^w\right)$ | Diameter is the maximum value of eccentricity ($e_i^w$). |

## 3. Description of the classification techniques

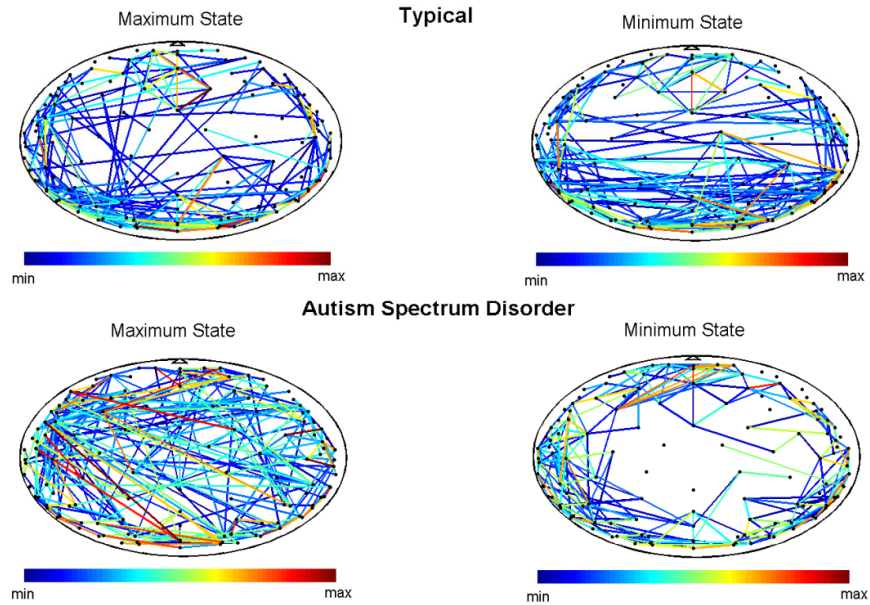

Figure 3: Brain connectivity for a typical and an ASD child.

Since the features have been extracted from a population of healthy and autistic children the problem is a binary classification task where the aim is to classify autism given the network measures of EEG phase synchronized states. The aim of this exploration is to find the optimal feature pool and classifier that can best distinguish between the two classes of subjects. Given this goal, the choice of the classifier is crucial for obtaining consistent classification results. The use of probabilistic classification approaches in the present context is not suitable due to the limited number of subjects, as it is not reliable to construct multi-dimensional probability densities functions (pdfs) from the features for Bayesian classifier and even one dimensional pdf for each feature in Naïve Bayes classifier [41]. Whereas, non-probabilistic classification techniques like discriminant analysis and SVM with polynomial kernels which map the feature vectors to a higher dimensional space in order to separate the classes using a linear separation boundary or hyper-plane can be used. However, their performance varies significantly depending on the assumption they make about utilizing all the data-points or the marginal data-points from two classes while adapting the classifier's weights in the training phase. Discriminant analysis gives emphasis to all the data points of the two classes to determine the weights of the classifier, thus is prone to get affected by outliers. In contrast SVM is based on the principle of maximizing the margin between the critical points (support vectors) of the classes.

### 3.1. *Discriminant analysis based classifiers*

The LDA classifier separates two classes using a linear decision boundary in the multi-dimensional feature space. The linear discriminant function is given by equation(4).

$$y = \sum_{i=1}^{N} x_i w_i + b = \sum_{i=1}^{N+1} x_i w_i = Xw \tag{4}$$

where, $y$ is the predicted class label ($y \in [-1,1]$), $N$ is the number of features, $x_i$ is the $i^{th}$ feature, $w_i$ are the weights and $b$ is the bias. Given an $N$ dimensional input, the corresponding decision boundary is given by a $(N-1)$ dimensional hyper-plane. If $y$ is greater than zero the object is assigned to one class and if $y$ is less than zero the input is assigned to the other class. A least squared estimation (LSE) based approach is commonly used to train the classifier's weight $w$ where the squared error of the predicted class and actual class is minimized. The classifier's optimum weight $w_{opt}$ is obtained in the form of pseudo-inverse of the input features $X$, multiplied by the class information vector $y$.

$$w_{opt} = \left(X^T X\right)^{-1} X^T y \tag{5}$$

LDA classifier performs well in data that is linearly separable. In practice especially in biomedical applications, more complex decision boundaries may be necessary. The use of higher order kernels is one way to circumvent this problem. Polynomial kernels transform feature vectors to a higher dimensional feature space. According to Cover's theorem any data-set can be made linearly separable in some higher dimensional space, if the order of the kernel is gradually increased [42]. The higher dimensional feature space can be created by performing nonlinear transformation on the input feature $x_i$ using a kernel function $k(x_i)$. For an example a polynomial kernel of order two (also known as Quadratic Discriminant analysis, in short QDA) produces a higher dimensional space with the original features plus their cross products. In the case of a two dimensional feature space with two-variable, $x_1$ and $x_2$, the quadratic kernel transforms the space into a 5-dimensional space with the variables $\{x_1, x_2, x_1 x_2, x_1^2, x_2^2\}$. Although the higher order kernels effectively increases the number of features by taking their inner products and use their combinations to train the classifier, the same least square technique is employed for discriminant analysis. However this increases the computational complexity and is prone to over-fitting resulting in failure to generalize on a new data-set, unless a large number of data-points are used in the training phase.

### 3.2. **Support vector machine (SVM)**

Contrary to the least-square approach for training discriminant analysis based (LDA or QDA) classifiers which give emphasis to all the data-points in the training-set while constructing the decision boundary, the SVM give priority to the critical data points that lie closest to the decision boundary and data-points of the other class. These critical points are known as support vectors. The classifier that maximize the distance between these critical vectors or support vectors are known as SVM. This approach is more likely to give a better separation of data as the basis lies on maximizing the margin between the support vectors producing the optimum hyper-plane. When the data is linearly separable in the original feature space, standard SVM uses a linear decision boundary. However, it tries to find a boundary which maximizes the margin ($M$) which involves using an optimization routine, with a constraint that all data points lie on the appropriate side of the hyper-plane. If the class labels are $y \in [-1,1]$ the decision boundary can be defined in between i.e. $y = 0$ following equation(6).

$$y = x_i w + b = 0 \qquad (6)$$

Given the value of $y$ at the support vectors must be $\pm 1$, we get $y_i(x_i w + b) \geq 1$, which means the optimization algorithm should yield $\{w, b\}$ which describes a hyper-plane in the feature space, such that the two classes fall on the appropriate side of the support vectors [42]. The margin $M$ can be derived as in (7).

$$M = \frac{(1-b)}{\|w\|} - \frac{(-1-b)}{\|w\|} = \frac{2}{\|w\|} \qquad (7)$$

The margin $M$, i.e. the distance between the lines separating the two classes is maximized by minimizing $\|w\|$. The minimization is constrained by the equation to ensure the boundaries are on the accurate side and is done using sequential minimal optimization (SMO), although the well-known quadratic programming (QP) can also be employed for the same purpose. When the data is not linearly separable, linear SVM is not that effective. In such cases, the data can be transformed into a higher dimension space using the kernel methods described earlier in section 3.2. However this is computationally intensive and is prone to over-fitting similar to the use of kernels in discriminant analysis. Higher the order of the kernel, the more complicated the decision boundary becomes and the chance of over-fitting increases. Although these complex boundaries may perform well on the training data, most of the time they fail to generalize with increasing order of the kernel. This particular phenomenon is observed since the classifier becomes prone to capture small inconsistent patterns underlying the data-set.

### *3.3. Cross-validation scheme to avoid over-fitting of classifiers*

A classifier should be able to generalize beyond the examples of the training set and the model should be able to make accurate predictions on unseen data. When the data available is small, in order to overcome the problem of over-fitting and reduce the sensitiveness to the choice of the training set a cross-validation technique allows more efficient use of the limited data. Given a data-set with $N$ observations, in the Leave-One-Out Cross Validation (LOOCV), each observation of the data is held out in turn for validating the model which is trained on the remaining $(N-1)$ number of data-points. Averaging over the resulting accuracies of $N$ time independent runs of the classifier training, gives the final average classifier accuracy. This ensures that all the test labels are shuffled and results in the best average estimate of the classifier accuracy. The LOOCV does not suffer from the problem of labelling which all other *N*-fold cross validation schemes possess. Since the LOOCV is the most extreme case of *N*-fold cross validation with *N* set to the number of data-points available, the chance of introducing an undesired bias is minimum [43]. Although the LOOCV is known to be computationally heavier than the *N*-fold cross validation scheme, for relatively manageable length of data-set it is preferred over the others. The use of the commonly used 10-fold cross-validation was restricted in this exploration by the limited number of subjects.

### *3.4. Preprocessing of Features, Feature Ranking and Classification Performance Measures*

In the machine learning literatures [44], there are two different paradigms of feature selection *viz*. scalar feature selection and feature vector selection. The scalar feature selection is independent of the classifier where the features are ranked using a score like Fisher's Discriminant Ratio (FDR) etc. For feature vector selection there are several suboptimal search techniques e.g. sequential forward search (SFS) and sequential backward search (SBS). Using a class separability criterion (like FDR) the poorly ranked features can be eliminated one by one or the best ranked features can be added to the feature subset to check a particular classifier's performance. Compared to the exhaustive search method, the suboptimal search techniques like FDR based feature ranking and grouping and adopting the SFS using these groups, is computationally less expensive. There could be several other possibilities of optimal feature selection considering dependency of the features, but the ultimate goal is to get a reliable and good classification.

Normalizing the features can remove bias from features having high value when training the classifier. Normalization scales the feature vector so that they are within the maximum ($x_{max}$) and minimum ($x_{min}$) value.

$$x_{norm} = \frac{x - x_{min}}{x_{max} - x_{min}} \tag{8}$$

The FDR is an efficient measure that allows finding the discriminating power of a feature and helps in dimension reduction. The larger the squared difference of the means of the features along with a small within-class variance, the better discriminating power the feature has. The features with higher FDR will have higher ranking implying they are compact and located distantly. The FDR of a feature is calculated using the mean and variance of individual classes i.e. $\{\mu_1, \mu_2\}$ and $\{\sigma_1^2, \sigma_2^2\}$ as (9).

$$FDR = \frac{(\mu_1 - \mu_2)^2}{(\sigma_1^2 + \sigma_2^2)} \tag{9}$$

We assess the classifier's performance as using the conventional measures sensitivity, specificity and accuracy. Typically the diseased class or abnormal condition is called, positive ($P$) and the typical or normal class as negative ($N$). The correct detection or true classification of abnormal conditions is known as true positive ($TP$). Likewise correct classification of typical population is true negative ($TN$). An incorrect classification can be of two types: classifying diseased as typical i.e. false negative ($FN$) and classifying typical as diseased i.e. false positive ($FP$). Sensitivity and specificity are also known as true positive rate ($TPR$) and true negative rate ($TNR$). These measures along with accuracy rate ($ACC$) are given by equation(10).

$$\text{Sensitivity or TPR} = \frac{\text{TP}}{\text{TP+FN}} \times 100\% = \frac{\text{TP}}{\text{P}} \times 100\%$$
$$\text{Specificity or TNR} = \frac{\text{TN}}{\text{TN+FP}} \times 100\% = \frac{\text{TN}}{\text{N}} \times 100\% \qquad (10)$$
$$\text{Accuracy or ACC} = \frac{\text{TP+TN}}{\text{TP+TN+FP+FN}} \times 100\% = \frac{\text{TP+TN}}{\text{P+N}} \times 100\%$$

## *4.* Results and discussion

One of the aims of this study is to find the optimal pool of features that can best distinguish the two classes of subjects. Collectively the data generates 36 features from the six network parameters ($N_{parameter} = 6$) i.e. corresponding to three stimuli (fear, happy and neutral i.e. $N_{stimuli} = 3$) with maximum and minimum occurring state ($N_{state} = 2$). As a result, their combination has yielded $N_{parameter} \times N_{stimuli} \times N_{state} = 36$ possible set of features for classification of the ASD from typical. Therefore, it might be interesting to look at which network parameter (in Table 1) or state (among max or min) or their combination has the best discriminating capability. Also, from the better discrimination point of view of ASD, the preference of the nature of stimuli (happy, angry and fearful face) can be analyzed from the FDR rankings.

The whole 36 feature set was broken down into nine different cases as shown in Table 2 so that we could determine the feature set that is most effective in classifying the data. In the first case (case-1) we used all the 36 features resulting from the combination of all the six network parameters corresponding to the maximum and minimum states for all the stimuli. This case allows us to find the best combination of network parameter, stimulus and max/min state which has the best discriminating power. For case-2 and case-3, only the features from the max states and min states were chosen respectively. From the results of case-2 and case-3 we may conclude which of the max state features or min state features are more efficient and even which state among the max/min is most effective in our current classification problem among the available $N_{parameter} \times N_{stimuli} = 18$ features in each case. For exploring the discriminative nature of the features (in each of the three cases 1-3) to separate the typical and autistic subjects, FDR is used to assign a ranking according to their decreasing order of importance. The FDR ranking and the values are shown in Figure 4-Figure 6 for the three cases respectively where the x-axis denotes the considered features and the relative weightage (FDR) is plotted in the y-axis. The ranked features are plotted in Figure 7 with decreasing order of importance using the FDR criterion. From Figure 7 it is evident that there exist four feature groups for case-1 to case-2 and three feature groups for case-3 where the features contained in one group have closer class-discrimination capability i.e. projections on y-axis are closer for the features in a single group. In case-1, the first group consists of top two features and next groups with three, four and 36 features respectively. In case-2, the four groups have top 2, 4, 7 and 18 respectively. In the case of all minimum-state features (case-3) the three groups have the top 7, then top 15 and then all 18 features as evident from Figure 7. Classification performance using LOOCV and different classifiers with these groups of FDR based ranking are compared next.

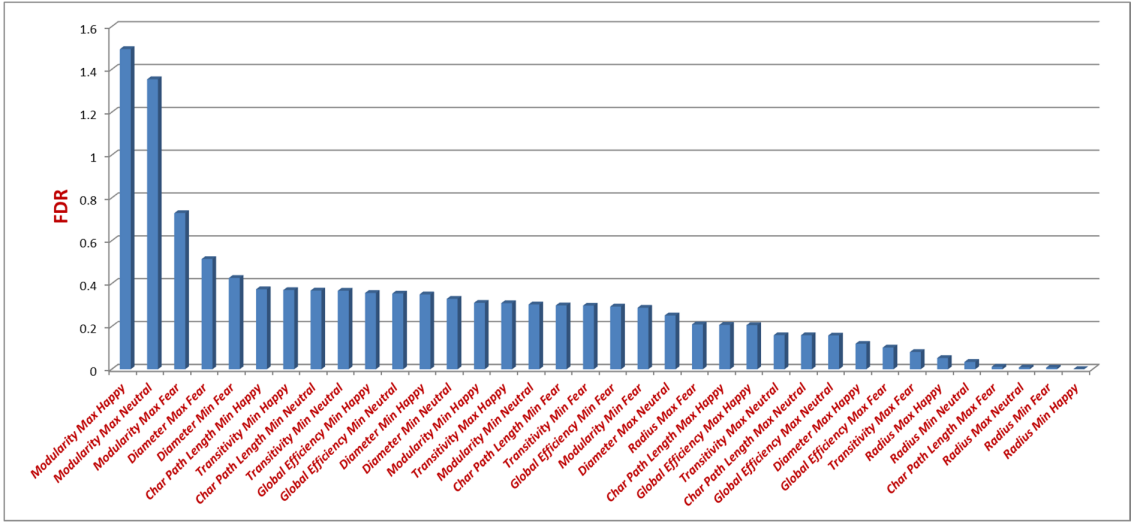

Figure 4: FDR ranking of different features for case-1 (all features).

Table 2: Different cases for classification considering max/min states, 6 network parameters and 3 stimuli

| Case number | Feature Combinations | Number of features |
|---|---|---|
| Case1 | All max and min state features for all 3 stimuli | 36 |
| Case2 | All max state features for all 3 stimuli | 18 |
| Case3 | All min state features for all 3 stimuli | 18 |
| Case4 | Transitivity for all 3 stimuli | 6 |
| Case5 | Modularity for all 3 stimuli | 6 |
| Case6 | Characteristic path length for all 3 stimuli | 6 |
| Case7 | Global efficiency for all 3 stimuli | 6 |
| Case8 | Diameter for all 3 stimuli | 6 |
| Case9 | Radius for all 3 stimuli | 6 |

The LOOCV classification performance for case-1 i.e. the entire feature set is shown in Figure 8. It shows that for the discriminant analysis increasing the number of features reduce the accuracy rate. This is in conjunction with our intuition that more features trained will cause over-fitting. When SVM was run on the data, we can notice that beyond SVM kernel order-2, the performance and the generalizing capability of the classifier reduces. The best accuracy for case-1 is 94.7% (with 85.7% sensitivity and 100% specificity) when the top 4 features are used to train an SVM classifier with a second order polynomial kernel. The top four features are the modularity values of the maximum states of all the three stimuli and the maximum state diameter for fearful face stimulus. This result is in agreement with the findings from [24] where the authors found modularity of synchrostates can distinguish between autistic and non-autistic classes.

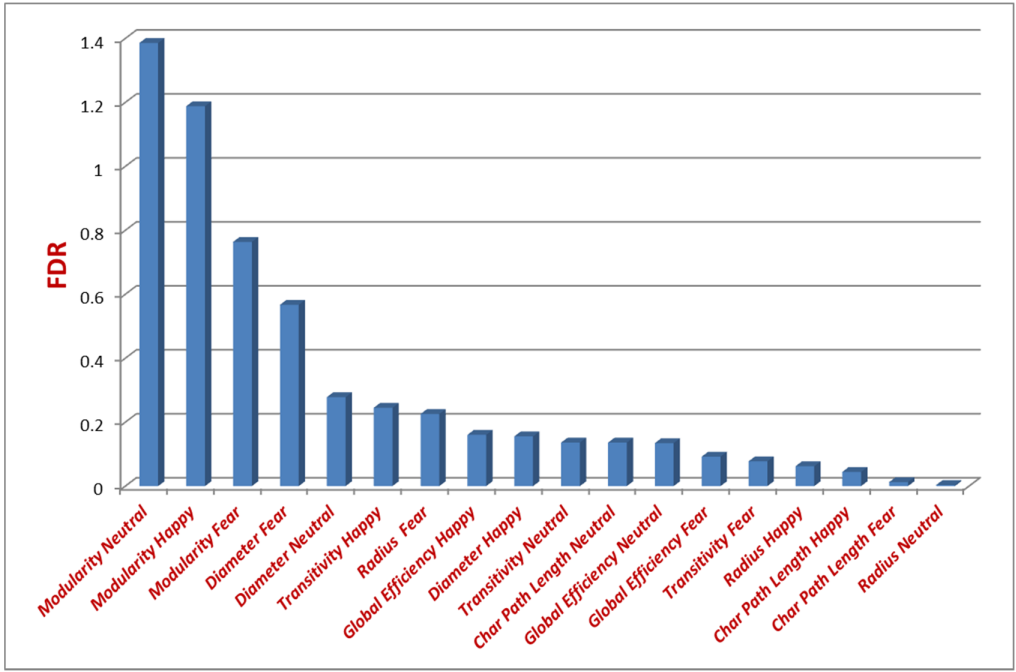

Figure 5: FDR ranking of different features for case-2 (max-state features).

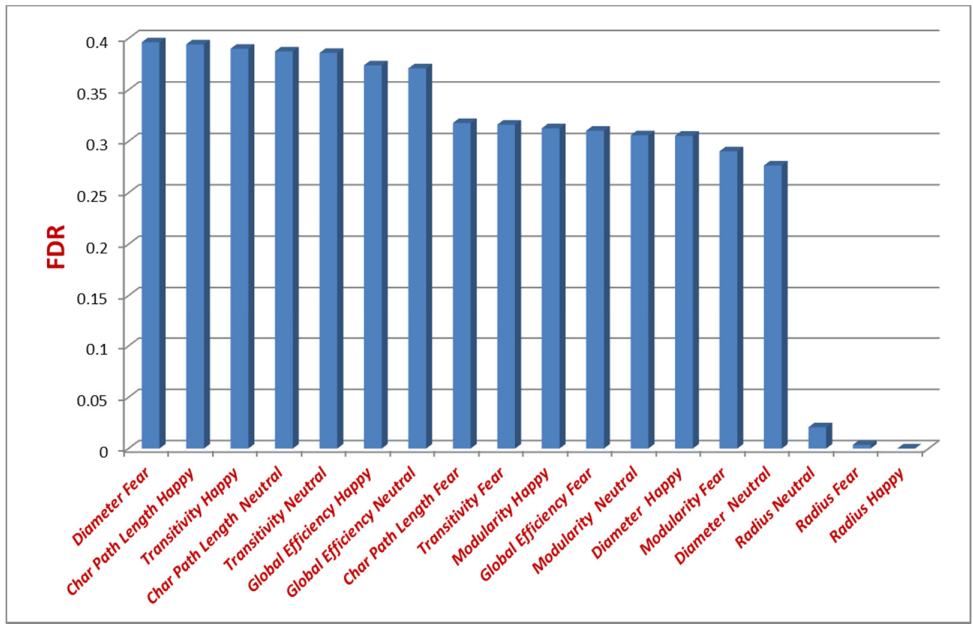

Figure 6: FDR ranking of different features for case-3 (min-state features).

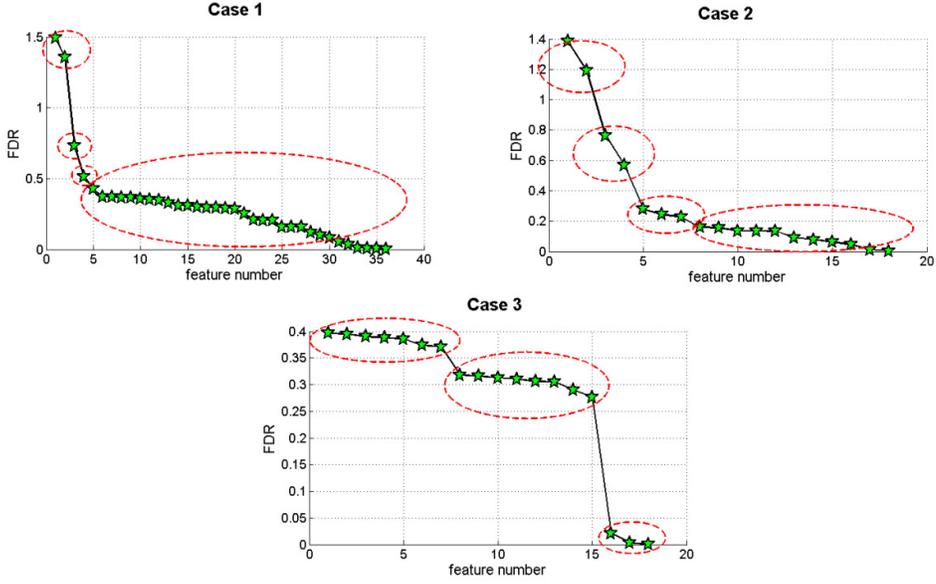

Figure 7: FDR based feature grouping for cases-1 to case-3.

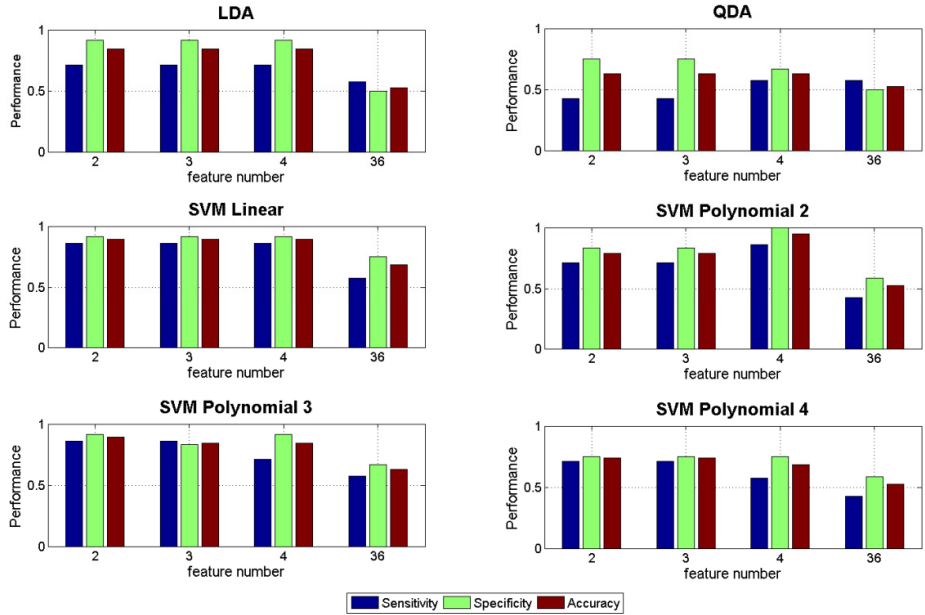

Figure 8: Performance of different classifiers with different group of features for case-1.

In case-2, we use only the maximum state features for all three stimuli for classification (Figure 9). We see that overall performance of QDA is poor compared to LDA. However when we use SVM with a linear and $2^{nd}$ order polynomial kernel our results are significantly better when compared to the discriminant analysis. Thus giving priority to the support vectors allows enhanced class separation. In this scenario, the highest accuracy value achieved is 94.7% (with 85.7% sensitivity and 100% specificity) which is the same for case-1 and so is the classifier configuration. The top four feature group contains the same features as of case-1.

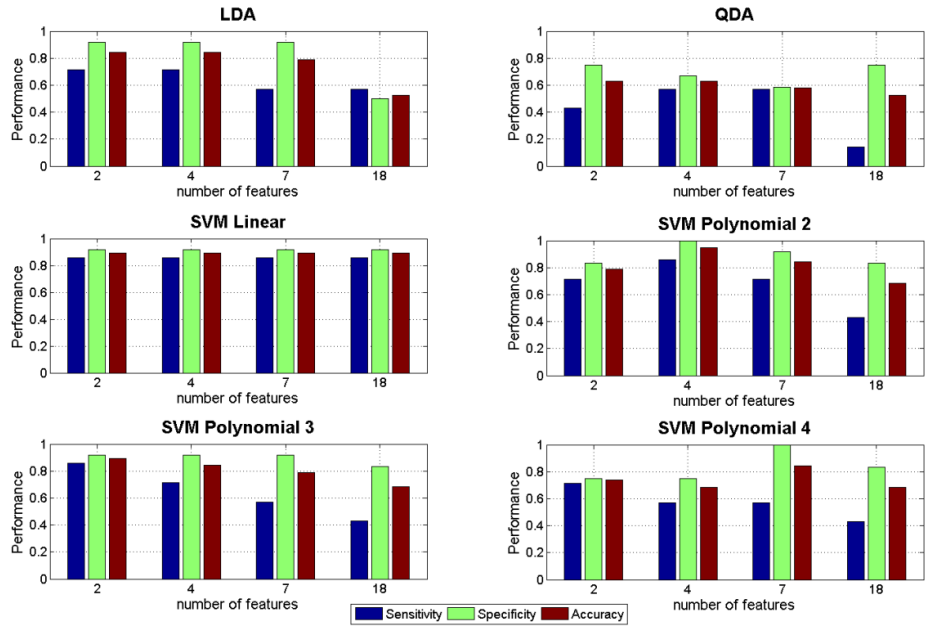

Figure 9: Performance of different classifiers with different group of features for case-2.

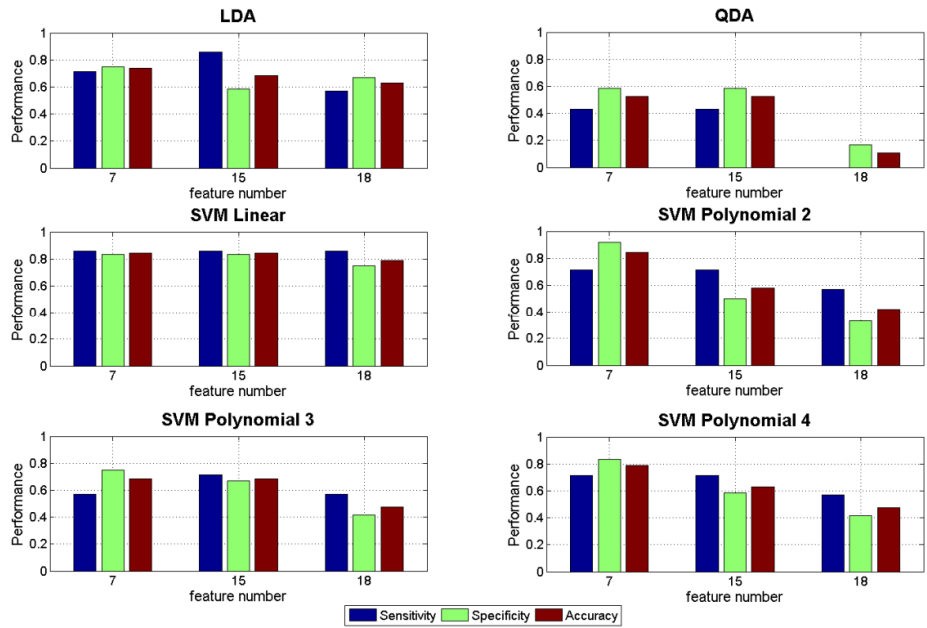

Figure 10: Performance of different classifiers with different group of features for case-3.

In case-3 shown in Figure 10, the overall accuracy levels are lower than compared to case-1 and case-2. The highest accuracy value achieved is 84.2% (with 85.7% sensitivity and 83.3% specificity). This is accomplished by applying SVM on the dataset in three cases i.e. with a linear kernel with top 7 features, with $2^{nd}$ order kernel on the top 7 features and a linear kernel with top 15 features. The configuration which has the least computational complexity is the linear SVM training with 7 features. Comparing this case with the previous two cases, we can say that the minimum state values are not as effective in differentiating between the two groups as the maximum state. There is also a disadvantage since more (seven) features need to be used to achieve this result in contrast to case-1 and case-2 where better accuracy can be achieved by using only 4 features.

Table 3: FDR based feature ranking for different network measures (case-4 to case-9)

| Rank | Transitivity (case4) | | Modularity (case5) | | Characteristic Path Length (case6) | | Global Efficiency (case7) | | Diameter (case8) | | Radius (case9) | |
|---|---|---|---|---|---|---|---|---|---|---|---|---|
| | FDR | Description | FDR | Description | FDR | Description | FDR | Description | FDR | Description | FDR | Description |
| 1 | 0.390 | Min Happy | 1.386 | Max Neutral | 0.394 | Min Happy | 0.374 | Min Happy | 0.567 | Max Fear | 0.226 | Max Fear |
| 2 | 0.386 | Min Neutral | 1.188 | Max Happy | 0.388 | Min Neutral | 0.371 | Min Neutral | 0.396 | Min Fear | 0.062 | Max Happy |
| 3 | 0.316 | Min Fear | 0.764 | Max Fear | 0.318 | Min Fear | 0.310 | Min Fear | 0.305 | Min Happy | 0.021 | Min Neutral |
| 4 | 0.245 | Max Happy | 0.313 | Min Happy | 0.136 | Max Neutral | 0.160 | Max Happy | 0.278 | Neutral Max | $3.63\times10^{-3}$ | Min Fear |
| 5 | 0.136 | Max Neutral | 0.306 | Min Neutral | 0.045 | Max Happy | 0.135 | Max Neutral | 0.277 | Neutral Min | $3.53\times10^{-3}$ | Max Neutral |
| 6 | 0.078 | Max Fear | 0.290 | Min Fear | 0.013 | Max Fear | 0.092 | Max Fear | 0.156 | Happy Max | $7.16\times10^{-5}$ | Min Happy |

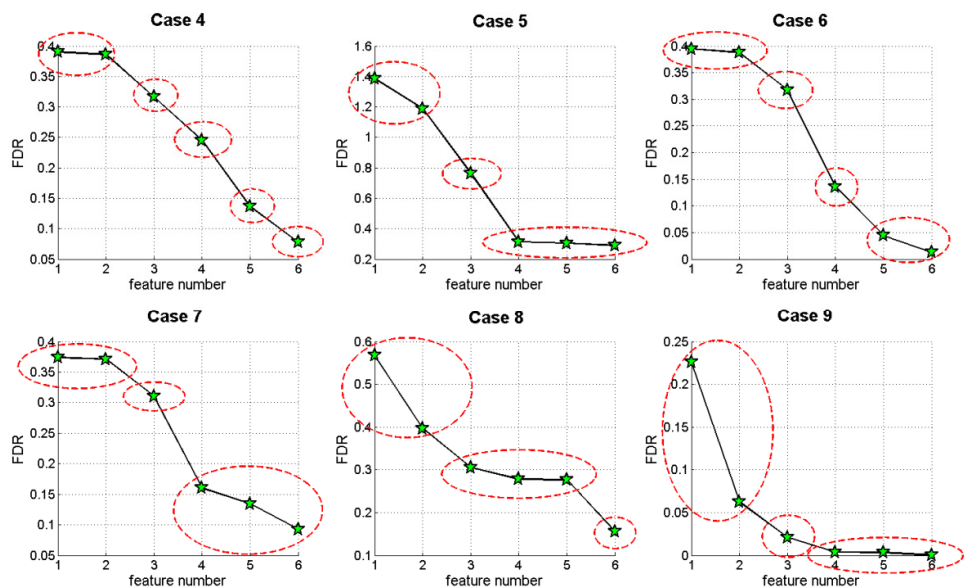

Figure 11: FDR based feature grouping for cases-4 to case-9.

The above case scenarios allowed us to factor out which combinations from the pool of features perform the best. It also gives an idea about the best classifier-setting that learns the dataset most effectively. The same principle is applied again but this time, to find out which of the complex network parameters have the best discriminating power. We design six cases with the maximum and minimum state values of the individual brain-network measures in Table 1 for all of the three stimuli. Hence each case has six features i.e. max/min state features for three stimuli for each choice of network parameters like transitivity, modularity, characteristic path length, global efficiency, diameter and radius. This will reveal the discerning capability of each of the complex network measures for three different stimuli i.e. happy, angry and fearful face. Each of the cases (from case-4 to case-9) has a pool of six features (i.e. $N_{stimuli} \times N_{state} = 6$) which represent the network metrics. In every case the FDR value was used to designate a rank to the feature in decreasing order of importance. The FDR value projection on the *y*-axis against the ranking number was used to group the features with the most discriminating power into one and so on. The FDR based feature grouping for the case-4 to case-9 has been elucidated in Table 3 and Figure 11. Although there are several algorithms available in the machine learning community for effective selection of least correlated features like scalar feature selection, sequential forward and backward selection etc., we here restricted our study with the FDR based feature grouping only, as it is a much simpler concept and easy to understand and implement. In fact, increasing the number of features using closely spaced FDR groups instead of individually adding them in the feature pool is quite similar to the concept of sequential forward feature selection method.

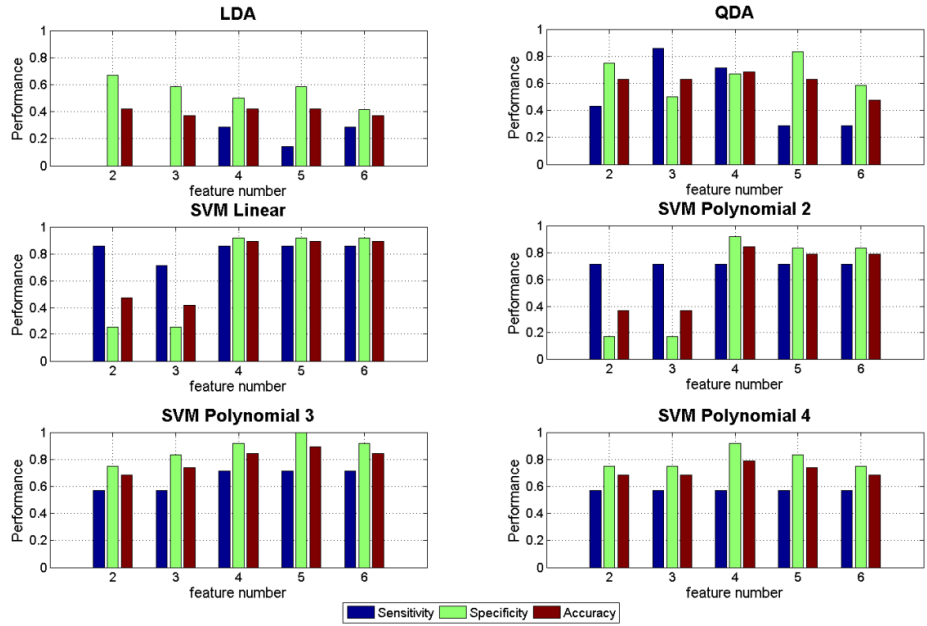

Figure 12: Performance of different classifiers with different group of features for case-4.

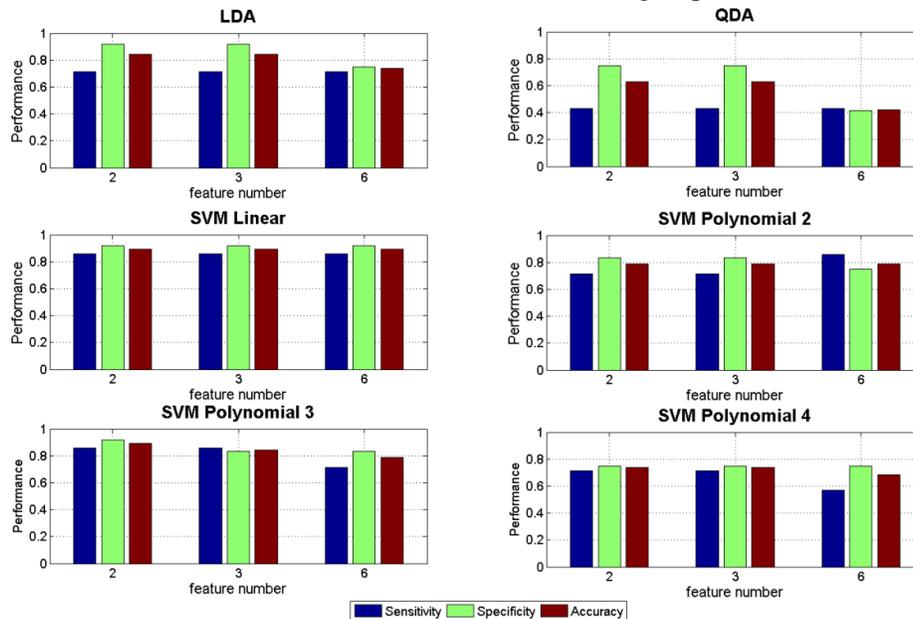

Figure 13: Performance of different classifiers with different group of features for case-5.

Case-4 used all the transitivity values for the maximum and minimum occurring states of all stimuli. Overall the discriminant analysis classifiers achieve poor results compared to SVM. When using SVMs it is noticeable that using more number of features is favorable in this case and yields better accuracy as evident in Figure 12. A top accuracy value of 89.5% (with 85.7% sensitivity and 91.7% specificity) is attained with a linear and 3$^{rd}$ order kernel of SVM using the top five features, without considering the maximum transitivity of fear (the lowest ranked feature). Training all six features with linear SVM also gives an accuracy of 89.5%. However, training the SVM using five features and a linear kernel will be most efficient due to less number of features.

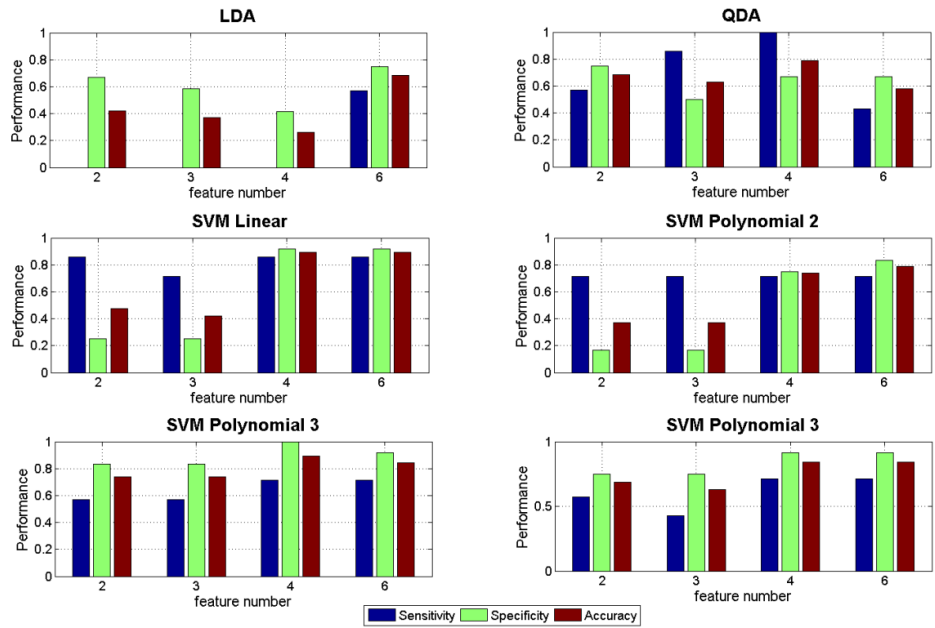

Figure 14: Performance of different classifiers with different group of features for case-6.

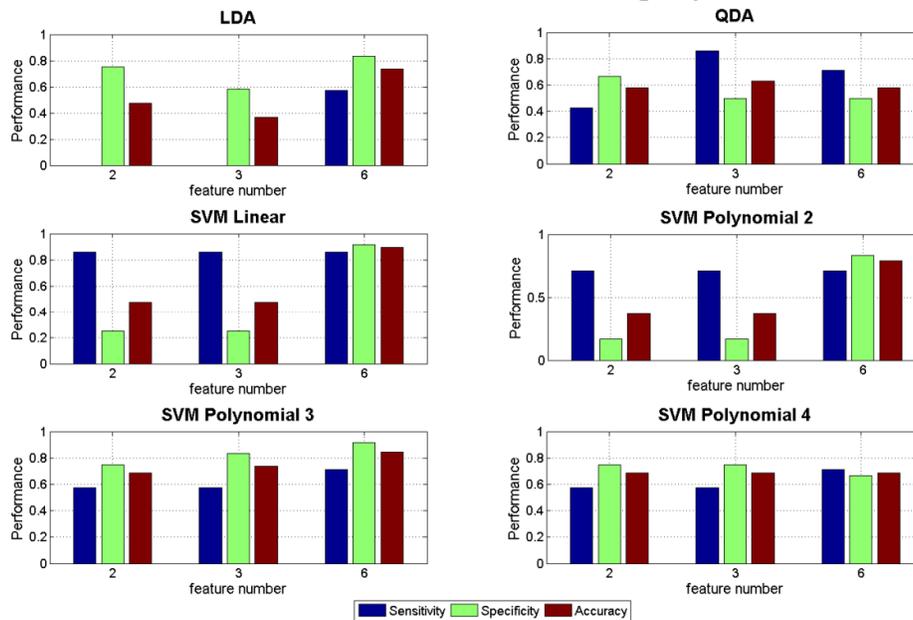

Figure 15: Performance of different classifiers with different group of features for case-7.

In case-5 where we consider all the modularity values we see similar grouping results with that of case-1 and case-2. The maximum modularity values are grouped as top 2 and top 3, whereas the modularity values for minimum state are grouped as least discriminant among all six features. The SVM classifiers have almost similar performance for different group of features while having the best accuracy for top two or top three features. Using all six features reduces the accuracy as it over-fits the data. Another interesting observation is increasing the SVM kernel order leads to poor performance as can be seen from Figure 13. From the SVM linear kernel plot, we observe, that accuracy reaches its maximum value of 89.5% (with 85.7% sensitivity and 91.7% specificity) for all feature groups. Increasing the number of features in this case is not improving the performance of the classifier. The maximum state modularity of the neutral and happy face stimuli when trained with a linear SVM is most effective in this scenario.

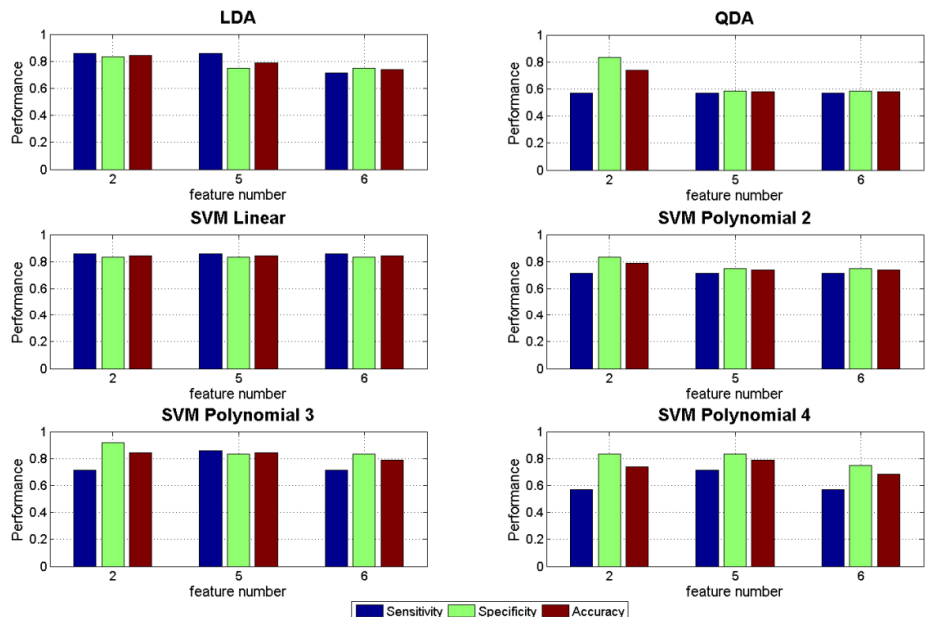

Figure 16: Performance of different classifiers with different group of features for case-8.

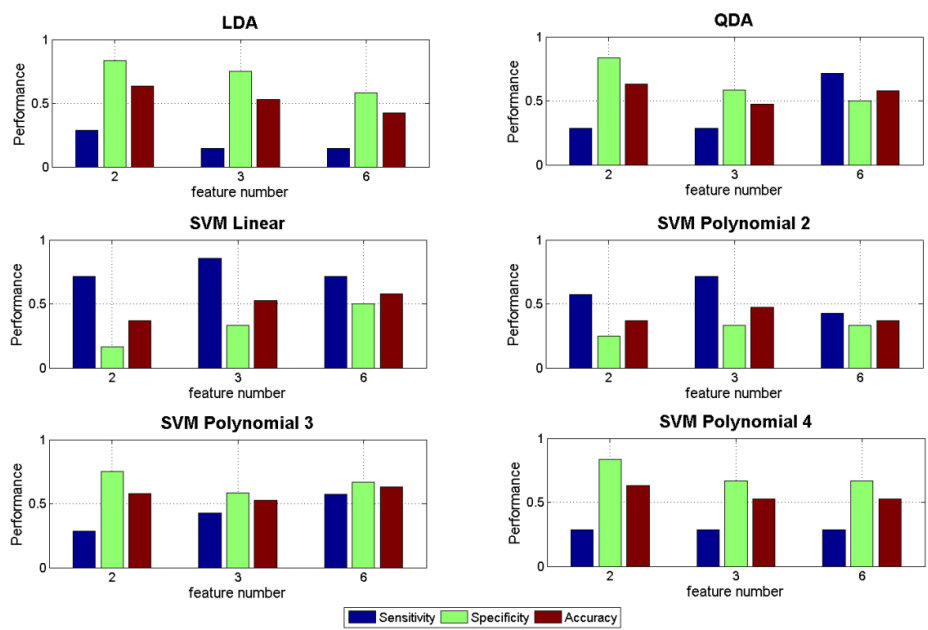

Figure 17: Performance of different classifiers with different group of features for case-9.

The classification results when considering only the characteristic path lengths (case-6) for all the three stimuli are given in Figure 14. The discriminant analysis techniques give poor performance for the characteristic path length features. In this case using more features increases the accuracy and the overall performance. This could be due to the quality of the feature. The classifier that achieves the best result is the linear SVM when applied to learn all six or top four features; however the latter requires less computational power. The best accuracy obtained was 89.5% (with 85.7% sensitivity and 91.7% specificity).

The overall performance of global efficiency features (case-7) is quite modest as shown in Figure 15. The accuracy levels with top 2 and 3 features are low for most of the classifiers. Using all features increases the accuracy with a maximum value of 89.5% (with 85.7% sensitivity and 91.7% specificity) when the linear SVM is trained using all the six features.

In case-8 we use all the diameter metrics as the features for classification. From the FDR plots the features are grouped as top 2, 5 and 6. Here the performance of LDA and linear SVM is comparable as shown in Figure 16. The best accuracy of 84.2% (with 85.7% sensitivity and 83.3% specificity) can be achieved by training the classifier with top 2 features with LDA and linear SVM. Using the top five features to train SVM with linear and $3^{rd}$ order polynomial also yields the same results. However LDA is much less complex than SVM training.

The overall performance of the network metric radius (case-9) is the worst among all the other complex network measures which is depicted in Figure 17. It can be concluded that this has the least discerning power for the current classification problem. None of the classifiers' accuracy in this case is satisfactory. The highest achievable accuracy of 63.2% (with 57.1% sensitivity and 66.6% specificity) is obtained using SVM with a $3^{rd}$ order polynomial kernel and all the features. This summarizes the exhaustive classification results of ASD vs. typical children with 6 network measures, two states and three stimuli while highlighting the best achievable accuracy with a particular classifier setting amongst the discriminant analysis and support vector family.

The summary of the key finding of the present paper for detection of ASD from graph theoretic measures from multi-channel EEG are as follow:

- The phase synchronization patterns or synchrostates in multi-channel EEG has been investigated in autistic and typical children during a face perception task. The complex brain network parameters have been extracted from the functional connectivity graphs of maximum and minimum occurring synchrostates which have been further used to classify an autistic group from the typical with an accuracy rate of 94.7% with corresponding sensitivity and specificity values of 85.7% and 100%.
- From the comprehensive study we can see that in most cases, considering more features during the training phase causes the gross accuracy to fall [45]. Also, the increase in complexity of the kernel does not always enhance the performance of the classifier due to over-fitting of the underlying inconsistent patterns.
- As a whole the maximum occurring state metrics have better discriminating capability than the minimum occurring state metrics.
- The best features to use for the classification of autistic children from normal ones is the maximum state modularity values during fear, happy and neutral stimuli and maximum diameter during fear stimuli. These features when trained using a $2^{nd}$ order polynomial kernel with SVM produces the best overall accuracy.
- The best graph metrics for classification are neutral and happy stimuli maximum state modularity.

## 5. Discussion

Apart from our study there have been a few attempts to detect or classify autism from EEG/MEG using machine learning algorithms. Bosl *et al.* [13] classified infants with high risk of autism vs. control group using SVM, *k*-nearest neighbors (*k*-NN) and Naïve Bayes classifiers. The study was based on modified multiscale entropy (mMSE) as feature for different age groups which resulted in an overall accuracy of 80%. Pollonini *et al.* [46] used Granger causality of MEG signals to discriminate autistic and healthy population using graph theoretic measures as features and SVM classifier which resulted in 87.5% accuracy. Discriminant analysis and SVM based classifiers were adopted in Stahl *et al*. [47] using event related potential (ERP) data resulting in an accuracy of 64% for discriminating between groups of high risk and low risk of autism. Compared to the approaches mentioned above here we first extract the maximum and minimum occurring synchrostates and obtained the associated brain network parameters which are then fed into the discriminant and SVM classifiers to differentiate ASD and healthy subjects. In our study, the overall classification accuracy (94.7% with SVM and four network measures) outperformed that reported in the previous mentioned literatures. It has been reported that individuals with ASD have long range functional under-connectivity and they compensate for this trait by forming more dense local connections in the frontal and posterior brain regions [48]. Complex network measures such as modularity, transitivity, global efficiency and characteristic path length which have been used as features in this paper, can effectively capture the integration (global connectivity) and segregation (local connectivity) ability of brain functional networks [38]. That is why the impact of these brain network measures is extensively investigated for potential classification between the ASD and typical cases.

It is well-known that Autism is a broad spectrum of disorders and a simple binary test may not be sufficient to make clinical decisions about the presence of autism. However, any EEG based evaluation method that can make a distinction between the two populations will be able to facilitate the clinicians in their behavioral assessment and prognosis. The synchrostates and the corresponding network measures effectively characterize the underlying functional brain connectivity of the subjects and hence may possess some signature of the particular characteristics of the ASD children. In order to obtain markers for the detection of ASD from the observation of EEG, the present work can be considered as the first step where small number of brain network measures can discriminate between the ASD and healthy subjects. The synchrostates might represent only a single biomarker of a very complex and heterogeneous spectrum of conditions such as ASD that require more complex clinical and neurobiological evaluation. The perspective of using synchrostates as one of the many tools used for the diagnosis of ASD is promising, but it needs further evidence obtained on larger and additional series. However, the severity or degree of ASD can further be classified in future using a similar procedure while deriving network measures from EEG synchrostates especially from different subclasses of ASD patients. Further extension of the present work can be directed towards further classification of the degree of ASD as a multi-class classification problem by using some psychological or behavioral assessment score as a threshold or decision boundary between the degrees of severity of ASD.

In addition, the phase synchrony, derived from EEG signals recorded over the scalp, has been doubted as it is believed to be the results of spurious synchronization that can occur due to volume conduction [49]. Synchronization induced through volume conduction decays with increasing distance between electrodes on the scalp [50]. The results presented in the brain connectivity diagram shown in Figure 3 shows that for the most part the strong connections are between distant electrodes and thus cannot be manifested due to volume conduction. The desynchronization and resynchronization property of different electrode signals over time i.e. the transition between the states in ms order of "synchrostates", is another property that cannot be explained by volume conduction [51]. The synchrony caused by volume conduction would result in a constant synchronization configuration existing over the scalp throughout the recording time for all synchrostates. If the synchrony obtained was the consequence of volume conduction it cannot not account for the change in synchronization patterns during state changes, in both strength and between relative electrodes over time. Research has also established and emphasized that the phenomenon of phase synchrony over the scalp extends to dynamic brain mapping [51] and is essential for the study of neural basis of cognition.

The purpose of the present paper was to show that brain network parameters derived from the synchrostates of EEG acquired while the children were performing a face perception task can effectively classify ASD and typical children. We also explored which experimental paradigm yields the most distinguishable result in terms of the nature of face stimuli i.e. fear, happy and neutral. Additionally, the best obtained brain network measures or features and the role of minimum and maximum occurring synchrostates for discriminating ASD and healthy subjects with relatively less complex classifier and kernels have been investigated.

Although the classification results shown in the current work is promising it is to be noted that a more rigorous prospective study with a large cohort of patients may be required to unequivocally eliminate the possible effects of misclassification and to establish the clinical validity of the technique before the methodology is put into clinical practice.

## *6.* **Conclusion**

In this paper, we successfully classify two groups of children with autism and typical using their 128-channel EEG signals extracted during face processing tasks. Complex network parameters were used as features to design and compare discriminant analysis and support vector family of classifiers with a maximum achievable accuracy of 94.7% using four features and a second order polynomial kernel in SVM. The study also revealed that the maximum occurring synchrostate holds the best discerning information and its modularity index can be considered as a unique biomarker for the detection of autism. Future work may be directed towards applying the brain connectivity measures for large population of ASD and typical children in order to effectively bring it in regular clinical practices.

**Acknowledgement**

The work presented in this paper was supported by FP7 EU funded MICHELANGELO project, Grant Agreement #288241. URL: www.michelangelo-project.eu/. We are grateful to Prof. Filippo Muratori for his valuable suggestions on this paper.